\documentclass[epj,nopacs]{svjour}
\usepackage{latexsym}
\usepackage{graphics}

\newcommand{\gt}{\ifmmode>\else{$>$}\fi}
\newcommand{\lt}{\ifmmode<\else{$<$}\fi}
\newcommand{\Gtsim}{{\ {\lower.6ex\hbox{$\sim$}}\llap{\raise.4ex\hbox{\gt}}}\ }
\newcommand{\Ltsim}{{\ {\lower.6ex\hbox{$\sim$}}\llap{\raise.4ex\hbox{\lt}}}\ }

\newcommand{\AP}{Astrop. Phys.}
\newcommand{\ApJ}{Ap. J.}
\newcommand{\ARNPS}{Ann. Rev. Nucl. Part. Sci.}

\newcommand{\NIM}{Nucl. Inst. Meth.}
\newcommand{\NP}{Nucl. Phys.}
\newcommand{\PL}{Phys. Lett}
\newcommand{\PR}{Phys. Rev.}
\newcommand{\PRL}{Phys. Rev. Lett}

\begin{document}
\hugehead             

\title{Search for Nucleon Decays
       induced by GUT Magnetic Monopoles\protect\\ with the MACRO Experiment}
%
%
\date{Received: \today / Revised version: \today}
\author{ 
M.~Ambrosio$^{12}$, 
R.~Antolini$^{7}$, 
G.~Auriemma$^{14,a}$, 
D.~Bakari$^{2,17}$, 
A.~Baldini$^{13}$, 
G.~C.~Barbarino$^{12}$, 
B.~C.~Barish$^{4}$, 
G.~Battistoni$^{6,b}$, 
Y.~Becherini$^{2}$,
R.~Bellotti$^{1}$
% ...
\\
\footnotesize
1. Dipartimento di Fisica dell'Universit\`a  di Bari and INFN, 70126 Bari, Italy\\
2. Dipartimento di Fisica dell'Universit\`a  di Bologna and INFN, 40126 Bologna, Italy \\
3. Physics Department, Boston University, Boston, MA 02215, USA \\
4. California Institute of Technology, Pasadena, CA 91125, USA \\
5. Department of Physics, Drexel University, Philadelphia, PA 19104, USA \\
6. Laboratori Nazionali di Frascati dell'INFN, 00044 Frascati (Roma), Italy \\
7. Laboratori Nazionali del Gran Sasso dell'INFN, 67010 Assergi (L'Aquila), Italy \\
8. Depts. of Physics and of Astronomy, Indiana University, Bloomington, IN 47405, USA \\
9. Dipartimento di Fisica dell'Universit\`a  dell'Aquila and INFN, 67100 L'Aquila, Italy\\
10. Dipartimento di Fisica dell'Universit\`a  di Lecce and INFN, 73100 Lecce, Italy \\
11. Department of Physics, University of Michigan, Ann Arbor, MI 48109, USA \\
12. Dipartimento di Fisica dell'Universit\`a  di Napoli and INFN, 80125 Napoli, Italy \\
13. Dipartimento di Fisica dell'Universit\`a  di Pisa and INFN, 56010 Pisa, Italy \\
14. Dipartimento di Fisica dell'Universit\`a  di Roma "La Sapienza" and INFN, 00185 Roma, Italy \\
15. Physics Department, Texas A\&M University, College Station, TX 77843, USA \\
16. Dipartimento di Fisica Sperimentale dell'Universit\`a  di Torino and INFN, 10125 Torino, Italy \\
17. L.P.T.P, Faculty of Sciences, University Mohamed I, B.P. 524 Oujda, Morocco \\
18. Dipartimento di Fisica dell'Universit\`a  di Roma Tre and INFN Sezione Roma Tre, 00146 Roma, Italy \\
$a$ Also Universit\`a  della Basilicata, 85100 Potenza, Italy \\
$b$ Also INFN Milano, 20133 Milano, Italy \\
$c$ Also Istituto IASF/CNR, 40129 Bologna, Italy \\
$d$ Also Universit\`a  di Trieste and INFN, 34100 Trieste, Italy \\
$e$ Also Dipartimento di Energetica, Universit\`a  di Roma, 00185 Roma, Italy \\
$g$ Also Institute for Space Sciences, 76900 Bucharest, Romania \\
$h$ Macalester College, Dept. of Physics and Astr., St. Paul, MN 55105 \\
$i$ Also Department of Physics, Pennsylvania State University, University Park, PA 16801, USA \\
$l $Also Dipartimento di Fisica dell'Universit\`a  della Calabria, Rende (Cosenza), Italy \\
$n$ Also RPD, PINSTECH, P.O. Nilore, Islamabad, Pakistan \\
$o$ Also Dipartimento di Scienze e Tecnologie Avanzate, Universit\`a  del Piemonte Orientale, Alessandria, Italy \\
$p$ Also U. Minn. Duluth Physics Dept., Duluth, MN 55812 \\
$q$ Also Dept. of Physics, MIT, Cambridge, MA 02139 \\
$r$ Also Intervideo Inc., Torrance CA 90505 USA \\
$s$ Also Resonance Photonics, Markham, Ontario, Canada\\
$*$ Corresponding author: sitta@to.infn.it}
\abstract{
The interaction of a Grand Unification Magnetic Monopole with a nucleon can lead to a
barion--number violating process in which the nucleon decays into a lepton and
one or more mesons (catalysis of nucleon decay). \\
In this paper we report an experimental study of the effects of a catalysis
process in the MACRO detector. Using a dedicated analysis we obtain new
magnetic monopole (MM) flux upper limits at the level of $\sim 3\cdot
10^{-16}~cm^{-2}~s^{-1}~sr^{-1}$ for 
$1.1\cdot 10^{-4} \le |\beta| \le 5\cdot 10^{-3}$\/, based on the search for catalysis events in
the MACRO data. We also analyze the dependence of the MM flux limit on
the catalysis cross section.
%
} 
\maketitle
\section{Introduction}
The supermassive magnetic monopoles (MMs) predicted by the Grand Unified Theories of
the electroweak and strong interactions (GUTs) have a core in which the
original Grand Unification symmetry is restored \cite{preskill84}. The
interaction of a nucleon with the MM core can lead to a reaction in which
the nucleon decays (catalysis of nucleon decay). The cross section for this
process is of the order of the geometrical cross section of the MM core,
$\sigma \sim 10^{-56}~cm^2$ \cite{kolb90,bais83}, practically negligible. In
1982 Rubakov \cite{rubakov81,rubakov83} and Callan \cite{callan82,callan83}
were independently led to the conclusion that this catalysis process may
proceed via another mechanism and have a much higher cross section, of the same
order of magnitude of the strong interaction cross sections.

The Rubakov--Callan mechanism is completely different from the spontaneous
decay and so are the corresponding final states. The main decay channels are
$e^+ \pi^0$ and $\mu^+ K^0$ for the proton decay and $e^+ \pi^-$ and $\mu^+
K^-$ for the neutron decay. According to the theoretical calculations, pions
should always come with positrons, strange particles with muons, and there
should be no neutrinos in the final states \cite{bais83}.

The catalysis cross section takes the form \cite{bais83,callan83,ellis82}
\begin{equation}
\label{eq:sigmacat}
\sigma_{\Delta B \neq 0} = {{\sigma_0}\over{\beta}}
\end{equation}
where $\beta$ is the MM velocity and $\sigma_0$ is a parameter, of the
order of the cross sections of the strong interactions, which depends on the
explicit QCD calculations. According to some authors \cite{rubakov83,arafune85}
however the cross section for proton decay has the form $\sigma \propto
1/\beta^2$, so a $1/\beta$ enhancement factor is expected in this case. In
addition there could be suppression factors for protons in a nucleus
\cite{arafune85}.

In the hypothesis of a non negligible cross section, the MM induced
nucleon decay can be exploited to detect the passage of a MM in an
apparatus. Many proton decay experiments (like Soudan \cite{bartelt87}, IMB
\cite{becker94} and Kamiokande \cite{kajita85}) and neutrino telescopes (such
as the Lake Baikal detector \cite{balkanov98}) have searched for mono\-poles
via the catalysis mechanism: a MM could have been identified as a series
of decay events more or less along a straight line. All these experiments have
established limits to the MM flux.

Beside these direct searches, bounds to the local mono\-pole flux can be set
by means of the catalysis process in an indirect way. Astrophysical objects
such as neutron stars, white dwarfs and also planets can accumulate MMs
by capturing the ones that impinge upon them \cite{kolb90}. In the presence of
nuclear matter a MM is able to release energy via the nucleon decay
catalysis. The limits are established by requiring the the energy released by
the catalysis cannot be greater than the observed one: in this way it is
possible to set an upper limit to the number of contained MMs and then to
translate it into a bound to the local MM flux
\cite{kolb90,kolb84,freese99}.

MACRO was a large scale multi--purpose detector located in the Gran Sasso
underground laboratory, optimized in the search for GUT magnetic monopoles with
velocity $\beta \ge 4\cdot 10^{-5}$ and with a sensitivity well below the
Parker bound as defined in \cite{parker82}. Direct searches for MMs were
made using various subdetectors both in a stand-alone and in a combined way in
different ranges of velocity. The final results on the upper limits to the
MM flux are given in Ref.\ \cite{monofin}; previous results can be found
also in \cite{ahlen94,ambrosio95,ambrosio97,ambrosio96,combined}. These limits
refer to a direct detection of bare magnetic monopoles of one unit Dirac charge
($g_D = 137/2e$), catalysis cross section $\sigma_{cat} < 1~mb$ and isotropic
flux.

In this paper we present a different approach to the search for magnetic
monopoles using the nucleon decay process. We developed a dedicated analysis
procedure aiming to detect nucleon decay events induced by the passage of a
GUT MM in the MACRO detector; the results of this search are reported as
a function of the MM velocity and the catalysis cross section. After
summarizing the properties of the MACRO streamer tube subsystem (Sect.\ 2),
the subdetector used in the present analysis, we describe the simulation
technique employed to study the characteristics of a catalysis process in the
detector (Sect.\ 3). The analysis criteria used in the search for these
particular events in the streamer tube data are reported in Sect.\ 4, and the
results of the real MACRO data analysis in Sect.\ 5. The previously published
MACRO MM search with the horizontal streamer tube subsystem is based on a
different analysis pattern in which the catalysis process is considered
negligible; in Sect.\ 6 we discuss how this analysis may be affected by the
induced nucleon decay and how the obtained upper limits to the MM flux
depend on the catalysis cross sections. The conclusions are given in Sect.\ 7.

\section{The MACRO detector}
MACRO had total dimensions of $76.5\times 12\times 9.3~m^3$\/, and consisted of
six ``supermodules'' arranged in a modular structure; each supermodule was
divided into a lower and an upper part, both equipped with streamer tubes and
liquid scintillation counters; in addition one horizontal and two vertical
planes of nuclear track detector were also present. A detailed description of
the apparatus can be found in Ref.s \cite{ahlen93,ambrosio01}. Special care was
taken to ensure that the three types of detectors and the readout electronics were sensitive to low $\beta$ particles. A single candidate event could have
provided distinctive and multiple signatures in the three subdetectors, so the
experiment had enough redundancy of information to attain unequivocal and
reliable interpretation on the basis of only few or even one event.

We looked for catalysis events in the data collected with the streamer tube
subsystem. Among the MACRO subdetectors the streamer tubes could be the most
sensitive to the catalysis process, having a sufficient combination of spatial
and temporal resolution to clearly separate the MM track from the
catalysis products. For the same reasons the streamer tube standard analyses
could be heavily affected by a possible catalysis event, therefore we studied
as well how the corresponding flux upper limit varies with the catalysis cross
section. Below we give more details on the streamer tube subsystem.
\begin{figure}
\resizebox{0.5\textwidth}{!}{
  \includegraphics{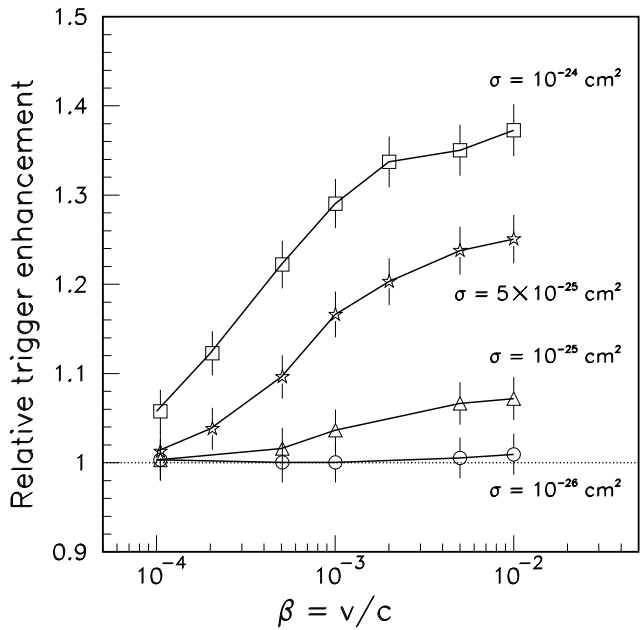}
}
\caption{Ratio of the total number of streamer MM triggers in a given
sample to the total number of streamer MM triggers in the sample without
catalysis.}
\label{triggers}
\end{figure}

The lower part of a MACRO supermodule was equipped with two horizontal planes
of scintillation counters and ten horizontal planes of streamer tubes
interleaved with seven rock absorber layers; the upper part was equipped with
one horizontal plane of scintillation counters and four horizontal planes of
streamer tubes without any absorber. The vertical west and east walls of the
apparatus were covered with six planes of vertical streamer tubes interleaved
with one vertical plane of scintillation counters; the frontal walls of the
lower part were covered with a similar structure. The tubes were filled with a
Helium--$n$-pentane mixture, in order to exploit the Drell--Penning effect
\cite{drell83,patera89} in detecting MMs, and operated in the
limited streamer mode. The cell dimensions and the charge drift in the gas
produced a time jitter, which resulted in a time resolution of about $140~ns$.

The streamer tube electronics were designed to be effective in the search for
slow charged particles with $\beta = v/c$ down to $10^{-4}$\/. Both the analog
and digital part of the tube signal were recorded. The analog part was used to
measure the time and charge of the streamer hits by means of a dedicated
circuit, the QTP system \cite{ahlen93,ambrosio01}, with a time resolution of
$150~ns$ and a charge resolution of $\sim 5\%$ the typical charge released by a
minimum ionizing particle. Charge and time information was kept in a cyclic
$640~\mu s$ deep memory. The digital part was used to determine the spatial
coordinate along a direction perpendicular to the tube anode wires, with a
resolution of about $1.1~cm$\/; stereo pick-up strips were used to determine a
second independent spatial coordinate with a resolution of about $1.2~cm$\/.
Combining these informations it is possible to reconstruct the particle track
in two space views and in a time view.

Two independent MM triggers employing the stre\-amer tube signals were
implemented, one with the horizontal planes of the lower part and one with the
vertical planes. The only assumption made in designing the triggers was that a
heavy slow-moving particle can cross the detector without any apparent
variation of direction or speed (which is the case of GUT MMs). In this
work only the trigger on the horizontal planes was used. The primary function
of this trigger was to measure the time-of-flight of particles in the range
$3\cdot 10^{-5} \le \beta \le 1$ and distinguish them from random noise. The
digital OR of all wires belonging to one plane was summed to the corresponding
signals from the next module, and then fed to the trigger circuitry, which
acted as 320 delayed coincidences (called {\it $\beta$-slices}\/) representing
the time-of-flight from plane 10 to plane 1 with a resolution of $3~\mu s$\/.
The trigger fired when the signals from at least seven planes of two
consecutive modules were aligned in time.

A complete description of the streamer tube hardware and the corresponding
triggers can be found in Ref.s \cite{ambrosio95,ahlen93,ambrosio01,demitri95}.

\section{The simulation}
To study the expected characteristics of the catalysis events induced in
MACRO and their consequences on the data analysis and results, a detailed MonteCarlo simulation of the physical process in the detector was performed.

The simulation of the catalysis process was independent of MACRO (and actually
can be applied to any detector set-up). The total cross section for the
catalysis was left as a free parameter of the simulation; the decay channels
and the branching ratios were chosen according to Ref.s \cite{bais83,ellis82},
namely $n,p \rightarrow e^+ \pi$ and $n,p \rightarrow \mu^+ K$ with branching
ratios 90\% and 10\%, respectively.
\begin{figure}
\resizebox{0.5\textwidth}{!}{
  \includegraphics{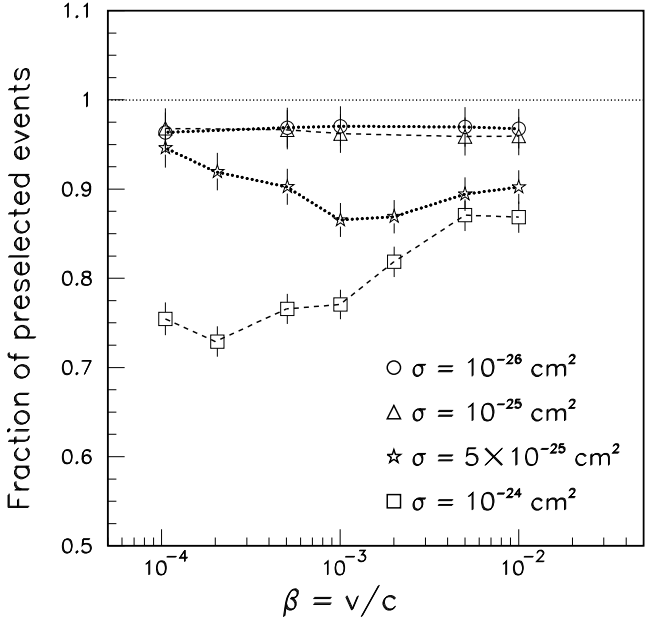}
}
\caption{Ratio of the number of selected events in a given sample to the number
of selected events in the sample without catalysis.}
\label{evenratio}
\end{figure}

Several improvements to GMACRO, the GEANT-ba\-sed simulation code of the
detector, were introduced. They concern a detailed description of the streamer
acquisition electronics, which took into account the hit time correlation when
multiple charges on wires and strips were recorded, the dead time of the
streamer acquisition chain, and the charge and time digitization performed by
the QTP system. It was inserted as well the simulation of the streamer tube
MM trigger on the horizontal wire planes, with a description of its
response to particles of different velocities and of its interaction with the
streamer data acquisition. In all these tasks we reproduced the behaviour of
the real electronic components.

Many samples of simulated events with different (and constant) MM
velocities and catalysis cross sections were simulated. Each sample consisted
of 10,000 single MM tracks generated from an isotropic flux. Five cross
section values were used, namely $\sigma_{cat} = 0, 10^{-26},$ $10^{-25},
5\cdot 10^{-25}$ and $10^{-24}~cm^2$ ($\sigma_{cat} = 0$ means no catalysis and
is used as a reference). For each cross section 5 (7 for $5\cdot 10^{-25}$ and
$10^{-24}~cm^2$\/) MM velocities were simulated: $\beta = 10^{-2}, 5\cdot
10^{-3}, 10^{-3}, 5\cdot 10^{-4}$ and $10^{-4}$ (plus $2\cdot 10^{-3}$ and
$2\cdot 10^{-4}$ for the larger $\sigma$). We kept constant the cross section
on nucleon $\sigma$ and not the parameter $\sigma_0$ (see Eq.
\ref{eq:sigmacat}), due to the uncertainties on both the $\beta$ dependence and
the possible suppression factors on various nuclei. In this way our results are
independent of the theoretical models, and can be used directly to place limits
on the cross section parameters.

We did not include in our simulation the rock around the detector, unlike other
experiments which took into account the possibility of a premature trigger due
to a catalysis event in the surrounding rock. In fact, given its trigger logic,
the streamer tube MM trigger on the horizontal planes could not be
activated by the decay products coming from a catalysis event in the external
rock. However in case of a fast particle trigger, the trigger signal was
conveniently delayed to let any slow moving particle cross and leave the
detector, which was not put immediately in dead time \cite{demitri95}.
Therefore we considered only MM catalysed nucleon decays inside the
detector.

As mentioned in the Introduction, for the cross section of the induced proton
decay there are two theoretical models: it could be either proportional to
$1/\beta$ as the cross section of the neutron decay, or to $1/\beta^2$\/. In
the latter case the proton cross section is enhanced by a factor $1/\beta$ with
respect to the neutron decay. To take into account both models, for each value
of the total cross section $\sigma_{cat}$ and for each $\beta$ we performed two
simulations: in one of them the decay probabilities of a proton and of a
neutron were equal, in the other one the probability of a proton decay was
enhanced by a factor $1/\beta$ with respect to the probability of a neutron
decay. The results showed a very small difference in the MACRO response to the
two models, well inside the statistical errors. In the following we show the
results only for the model in which both cross sections are equal, the results
for the other model being very similar; when physical conclusions are drawn,
results for both of them are given.
\begin{figure*}
\resizebox{\textwidth}{!}{
  \includegraphics{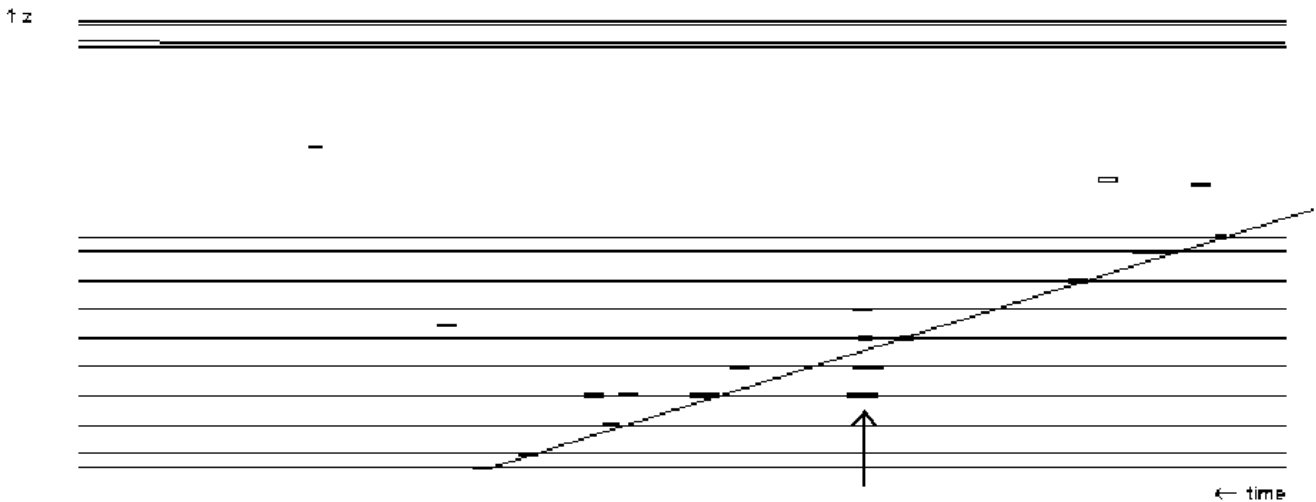}
}
\caption{Time view of a simulated MM induced nucleon decay inside the
detector: the reconstructed slow MM track and the fast particles hits (at
the arrow position) are clearly distinguishable.}
\label{eventtime}
\end{figure*}

From the experimental point of view it is interesting to notice how the
detector response changes due to the catalysis process. Fig.\ \ref{triggers}
shows the ratio of the total number of streamer MM triggers in a given
sample over the number of streamer triggers in the sample without catalysis
(i.e.\ with $\sigma_{cat} = 0$\/) for the same $\beta$ (in this Figure and in
all the subsequent ones the error bars are statistical, and smaller than the
marker size where they do not appear; only the samples at the marker position
were simulated and analyzed, the connecting lines serve only as a guide to the
eye). Clearly the number of triggers increases with $\sigma_{cat}$ and the
MM $\beta$\/, since the wire signals produced by the decay products can
contribute to the trigger formation. There can be even $\sim 40\%$ more
triggers for the highest values of $\sigma_{cat}$ and $\beta$ which were
simulated. The trigger increase is small at low velocities and larger at high
velocities, because the catalysis hits are very near to each other in time and
can easily contribute to activate the lower trigger $\beta$-slices (the ones
triggered by fast moving particles), but can rarely contribute to the higher
ones (interested by slow moving particles). For $\sigma = 10^{-24}~cm^2$ there
are so many catalyses per event that the secondary particles hits can activate
the trigger by themselves, determining the larger value of the trigger ratio
also for the lowest $\beta$ value.

\section{Search for nucleon decay events}
All the MonteCarlo samples were analyzed with exactly the same programs used
later on for the real data. Only the horizontal streamer planes of the lower
MACRO participated to the trigger formation, but for the event reconstruction
all horizontal planes belonging to both the lower and the upper parts of the
detector were used.

Events were preselected by requiring
\begin{enumerate}
\item at least one space track matching in both space views
\item each track must have at least 5 points in each view
\item in case of multiple wire tracks, at least one ``non--parallel'' track
\item at least one time track with $|\rm{ToF}| \ge 0.8~\mu s$
\end{enumerate}

Conditions 1 and 2 identify good MM tracks, the third one rejects
multiple muons from cosmic rays, which sometimes may confuse the time tracking.
The last condition selects slow moving particles (a time-of-flight of
$0.8~\mu s$ corresponds to a MM crossing vertically the ten lower
horizontal planes with $\beta = 2\cdot 10^{-2}$\/, a value well above the upper
validity limit of the streamer tube analyses).

In Fig.\ \ref{evenratio} the ratio of the number of selected events in a given
sample over the number of selected events in the sample without catalysis is
shown as a function of the MM $\beta$ and the catalysis cross section
$\sigma_{cat}$\/; this ratio is actually the relative efficiency with respect
to the case of no catalysis. The main reason of inefficiency is a wrong ToF
estimation because sometimes when there are many catalysis hits the time fit
may get confused and mix the MM signals with the secondary particle
ones, giving thus a time-of-flight lower than the actual value. Clearly the
cleaner the event the more correct the ToF computation, so for small cross
sections this effect is negligible, while it is much more relevant for large
cross sections. As a consequence these criteria result more efficient in
selecting low $\sigma_{cat}$ events, and even more efficient for no catalysis
events. Nevertheless these cuts, the ToF cut in particular, cannot be relaxed
without contaminating the preselected samples from real data with atmospheric
muons events (especially if accompanied by noise or showers).

By examining the simulated events it appears that the space view of a catalyzed
decay is indistinguishable from a cosmic muon accompanied by some noise, while
it has a typical pattern in the time view: a slow track due to the MM
with some ``fast'' hits, generated by the catalysis products, contained in a
small ($\Ltsim 1~\mu s$) time window. In Fig.\ \ref{eventtime} the time view of
a simulated catalysis event is shown: the slow track of the MM crossing
the detector and the hits of the fast decay particles (at the arrow position)
are clearly distinguishable . This peculiar signature can be exploited in a
catalysis--oriented analysis, which is applied to the events selected with the
preliminary cuts described above. The characteristics of the time view of a
catalysis event, and in particular the presence of fast hits grouped together
along an almost vertical track, are the main feature used in the present search
for catalyzed nucleon decays, while being a possible source of inefficiency for
the previously published MACRO direct MM searches with the streamer
tubes, which assumed no decay induced by the MM. Although the space view
cannot be used to tag an event as a catalysis candidate, nevertheless some
space consistency must be required anyway, for instance the catalysis hits
cannot be too far from the MM track.

Since the streamer trigger requires at least 7 planes, a track must consist of
at least 7 points. In the present analysis this condition had to be loosened,
since also the catalysis hits can contribute to the trigger formation, which
may be activated even by shorter tracks: so each track is required to consist
of at least 5 points only. If the MM traversed the detector, it must be
reconstructed also in the time view and be a slow particle, hence a $\beta$ cut
had to be imposed to the time track. There must be a consistency between the
wire and time views: the same planes must be present in the wire as in the time
fit, since they are both reconstructed from the signals coming from the same
hardware elements; possible hardware inefficiencies are taken into account by
leaving the possibility of 1 or 2 planes present in one fit and not in the
other. Catalysis hits are searched for in $\pm 10~\mu s$ around the time track,
because they cannot be too far in time from the MM. At least one
catalysis hit must be within $1~m$ around a wire track: the catalysis must have
been initiated in the vicinity of the MM; in this way we exclude random
noise hits uncorrelated in space but accidentally near in time. All catalysis
hits must be inside a $1~\mu s$ time window, since the decay products are fast
particles. Finally, catalysis hits must be present on at least 3 out of 4
consecutive planes: this is the minimum requirement needed to clearly identify
a catalysis candidate while rejecting the background.

\begin{figure}
\resizebox{0.5\textwidth}{!}{
  \includegraphics{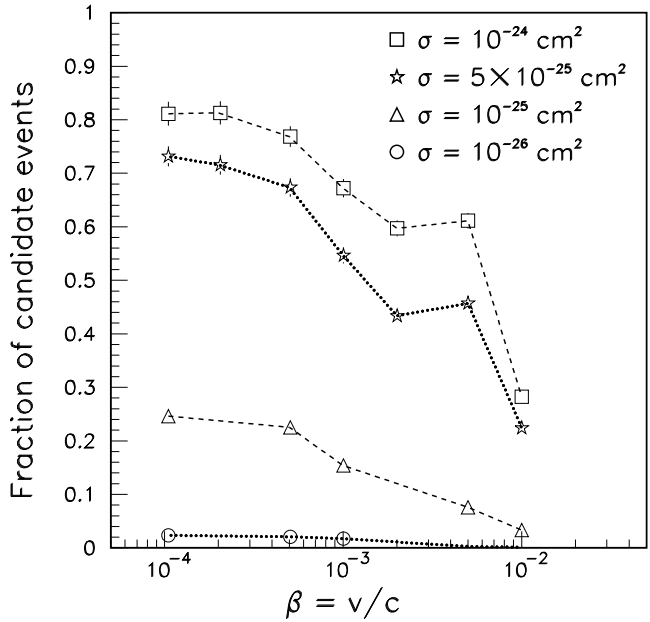}
}
\caption{Fraction of preselected simulated events with catalysis which pass the
catalysis specific cuts.}
\label{candratio}
\end{figure}

The above features have been implemented in the analysis of the preselected
events following these steps:
\begin{enumerate}
\item select an event with at least one track in both the spatial and temporal
views with at least 5 points
\item require no more than 2 planes fitted in either view and missing in the
other one
\item for each time track with $|\beta| \le 10^{-2}$\/, which is assumed to be
the MM track, define a time window equal to the time intersection of the
track with the first and tenth plane $\pm 10~\mu s$ in which to search for
catalysis hits
\item if hits on at least 3 out of 4 consecutive planes in a $1~\mu s$ window
are found, of which at least one within $1~m$ from any wire track and the other
ones in $\pm 2~m$ from the first one, keep the event as a catalysis candidate.
\end{enumerate}

Fig.\ \ref{candratio} shows the fraction of preselected simulated events with
catalysis which meet all the aforementioned conditions. For the samples at low
cross sections ($\sigma_{cat}$ $\Ltsim 10^{-25}$ $cm^2$\/) this analysis is
very inefficient, so our limits will be significant only for larger cross
sections. The strong decrease of the efficiency for $\beta = 10^{-2}$ is due to
the $\beta$ cut on the time track.

The same analysis scheme was applied also to the simulated sample without
catalysis. No event remained, so we can conclude that these cuts have a $100\%$
efficiency in rejecting simulated events without catalysis. As a further check,
all events without catalysis contained in the samples at lower cross sections
were rejected. 
\begin{figure*}
\resizebox{\textwidth}{!}{
  \includegraphics{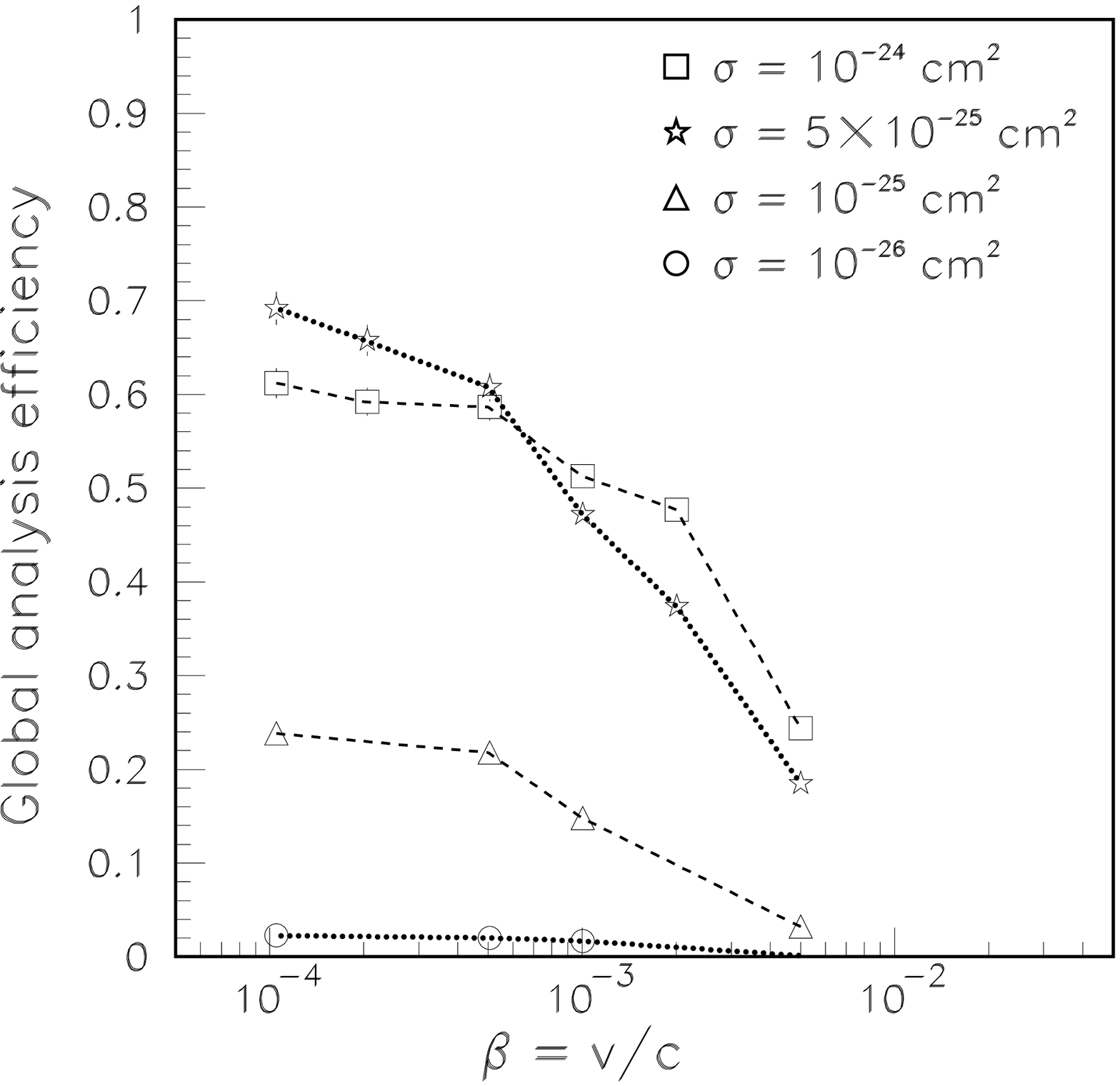}
  \includegraphics{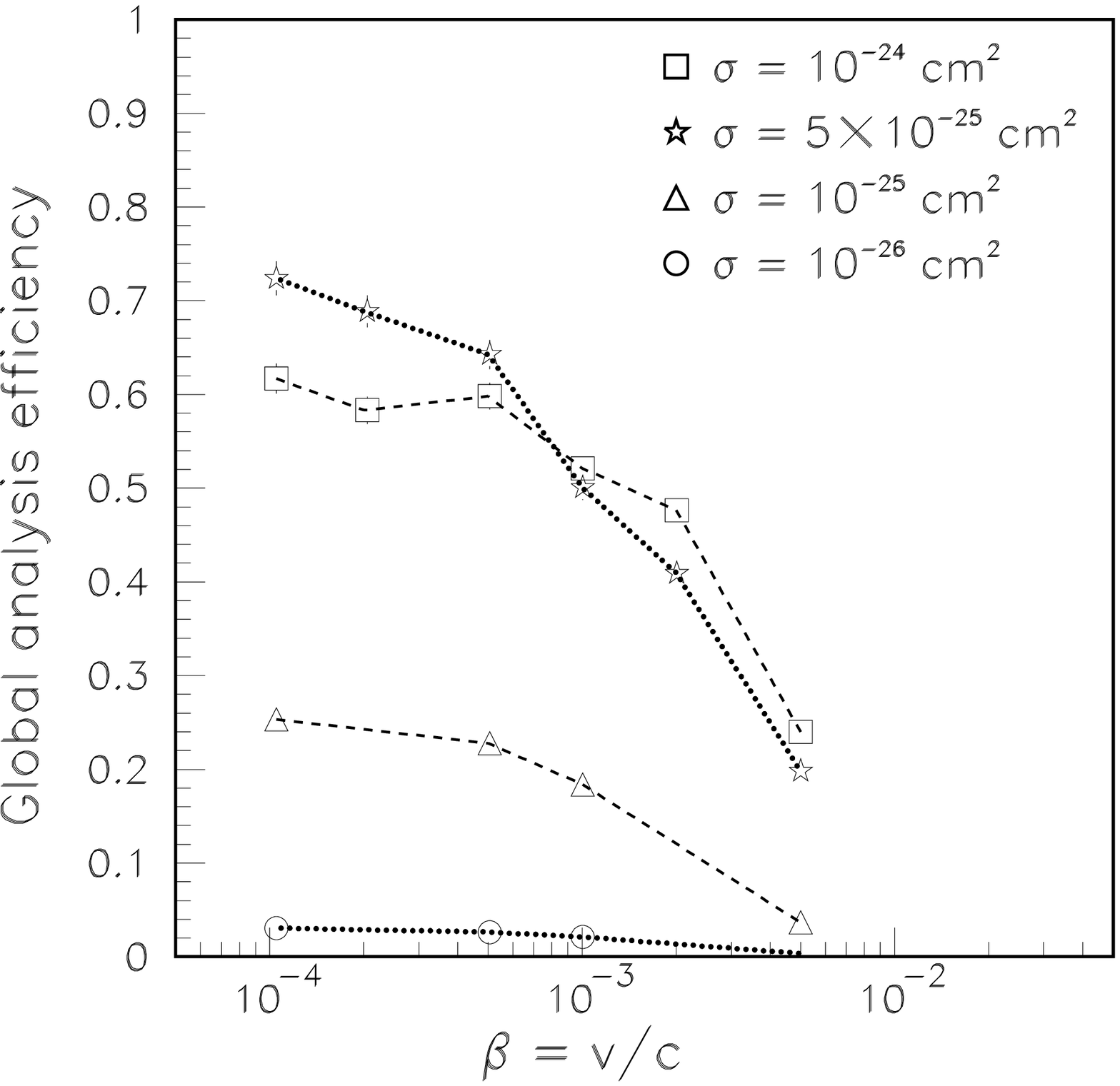}
}
\caption{Fraction of simulated events with catalysis identified by the analysis
with respect to the total number of events in each sample as a function of the
MM velocity for various cross section values (including the noise cut):
equal cross sections for proton and neutron decay on the left, enhanced proton
decay cross section on the right.}
\label{finaleff}
\end{figure*}

We required that the present catalysis search was valid in the range $1.1\cdot
10^{-4} \le |\beta| \le 5\cdot 10^{-3}$\/, which is the same range of the
streamer standard analysis: the lower limit is the minimum MM velocity
required for the Drell--Penning effect to take place, the upper limit is
required to fully reject the atmospheric muon contamination. In Fig.\
\ref{finaleff} the fraction of initial events (that is, events contained in the
MonteCarlo generated samples) which survive to all cuts (including the cut to
reject the electronic noise, described in the next paragraph) for both cross
section models is shown. This fraction can be considered the total efficiency
of the present data analysis (the efficiency of the hardware is assumed to be
the same of the published MACRO streamer tube analysis
\cite{monofin,ambrosio95}).

\section{Data analysis}
The same analysis cuts described in the previous paragraph were applied to the
whole sample of real data collected with the streamer tube MM trigger on
the horizontal planes, the same sample used also in \cite{monofin}. These data
were collected from January 1992 to September 2000, for a live time of 71193
hours; the full configuration detector acceptance, computed by means of a
MonteCarlo simulation including geometrical and trigger requirements, was
$4250~m^2~sr$\/.

From a visual scan of a small subsample of preselected real events, some of
them were found to have heavy electronic noise: typically there are two or more
horizontal planes in which all the QTP channels have fired; for this reason the
corresponding time hits are biased, so as to compromise the ToF measurement.
These events represent a clearly spurious background which the real samples are
rich of. In order to reject the noisy events, a further cut was imposed on the
preselected events before applying the catalysis identification analysis,
namely to have no more than 4 adjacent fired QTP channels. This additional
requirement had no effect on the simulated samples with lower cross sections,
and was at the level of few percent for higher cross sections. The effect of
this cut is already included in Fig.\ \ref{finaleff}\/.

\begin{table}[tb]
\begin{center}
\caption{Results of the real data analysis}
\begin{tabular}{r|l}
\hline
Cuts applied~~~~~~~~~~~~~ & Events survived \\
\hline
Total number of preselected events & 86339 \\
$\le 4$ adjacent fired QTP channels & 67565 \\
Consistency of wire and time tracks & 62130 \\
Catalysis specific cuts & \phantom{123}15 \\
validity range: $1.1\cdot 10^{-4} \le |\beta| \le 5\cdot 10^{-3}$ &
                          \phantom{1234}0 \\
\end{tabular}
\end{center}
\end{table}

\begin{figure*}
\resizebox{\textwidth}{!}{
  \includegraphics{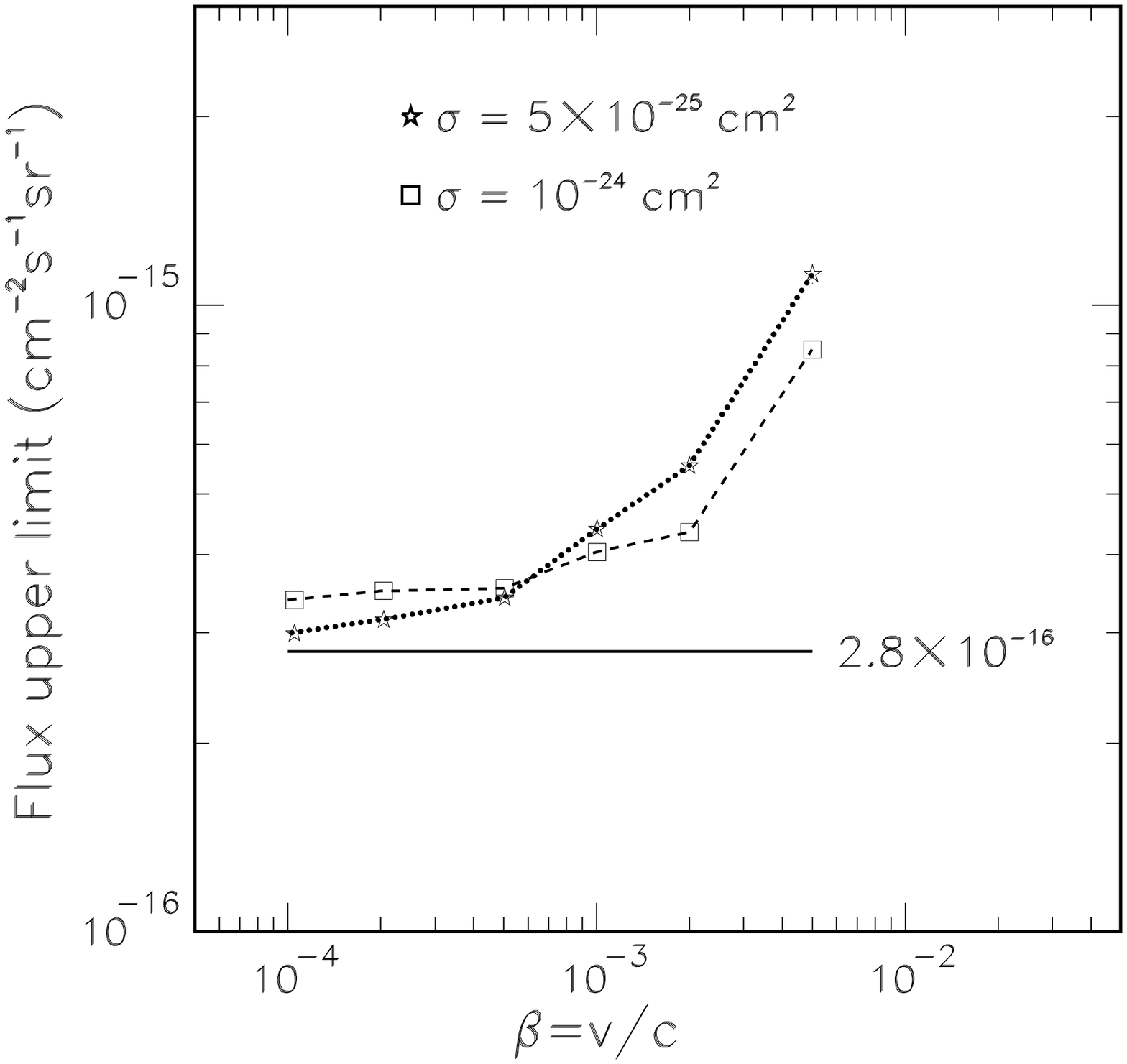}
  \includegraphics{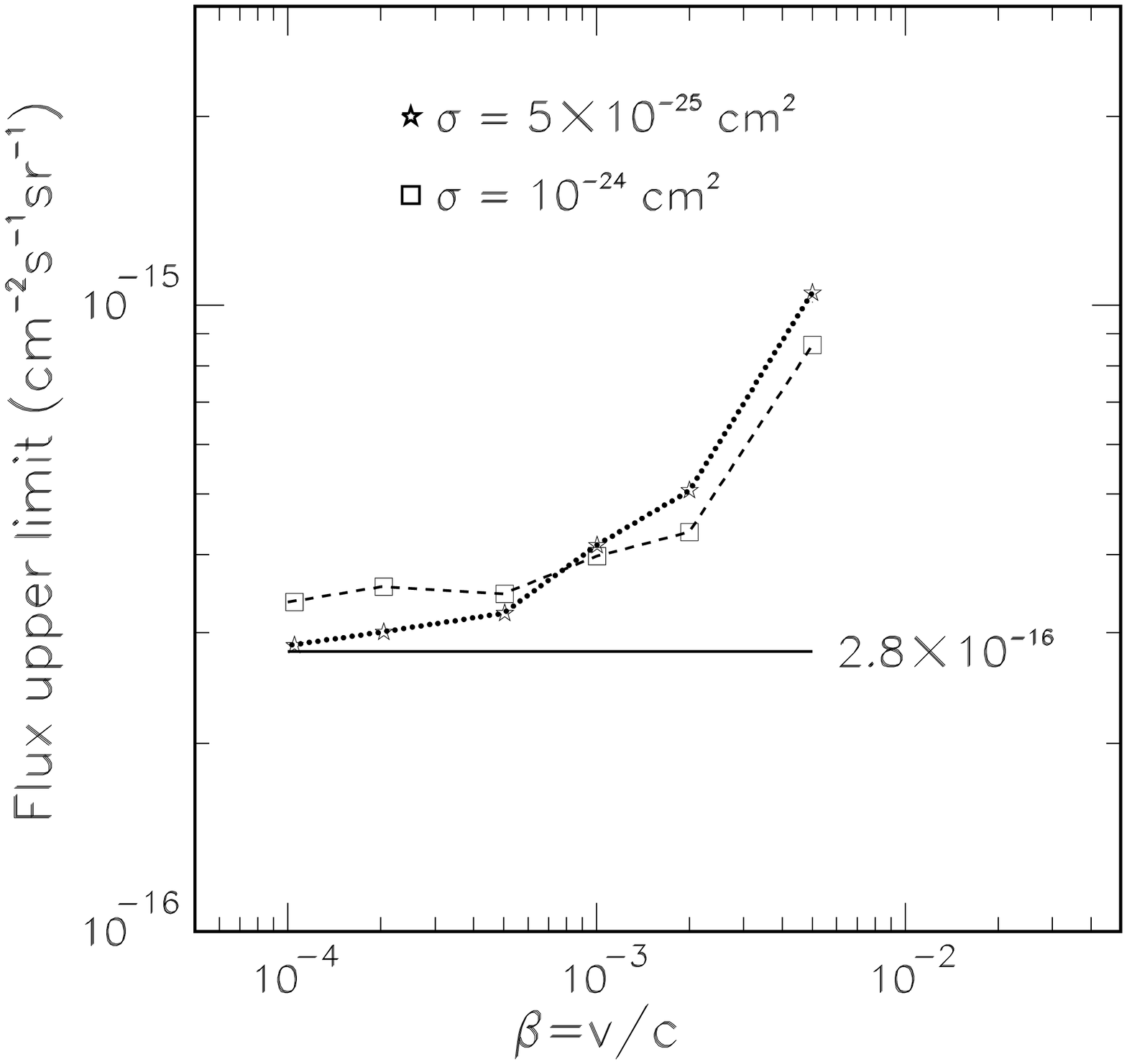}
}
\caption{Upper limits to the MM flux as a function of the MM
velocity for two values of the catalysis cross section;
in the left panel the proton and the neutron catalysis cross sections are equal, while in the right panel the proton cross section is enhanced by $1/\beta$, see text. The solid line is the standard limit for MM without catalysis, using the  streamer tube).}
\label{limits}
\end{figure*}

The events were preselected with the loose criteria described in the beginning
of the Section 4; to these events the catalysis-specific cuts were applied. In
Table 1 the number of events which remain after each cut is reported. No
candidate survived, therefore upper limits to the mono\-pole flux were set as a
function of the cross section. We limited this analysis to the larger values
since for $\sigma_{cat} \le 10^{-25}$ $cm^2$ the global efficiency is too low.
The time-integra\-ted acceptance was the same as for the ``standard'' streamer
tube analysis \cite{monofin,ambrosio95}, since exactly the same real data
sample was analyzed. 
The new upper limits are plotted as a function of $\beta$ in Fig.\ \ref{limits}:
the solid lines represent the standard streamer limit in the hypothesis
of no catalysis. Only the plotted points were actually computed, and the lines are only rough interpolations to guide the eye.
The lowest bounds at $\beta = 10^{-4}$ are $2.99$ and $3.38\cdot
10^{-16}~cm^{-2}~s^{-1}~sr^{-1}$ for $\sigma_{cat} = 5\cdot 10^{-25}~cm^2$ and
$10^{-24}~cm^2$ respectively, for the model with equal cross sections, and
$2.86$ and $3.36\cdot 10^{-16}~cm^{-2}~s^{-1}~sr^{-1}$ for the same cross
section values, for the model which assumes a proton cross section enhancement
by a factor $1/\beta$\/.

\begin{figure}
\resizebox{0.5\textwidth}{!}{
  \includegraphics{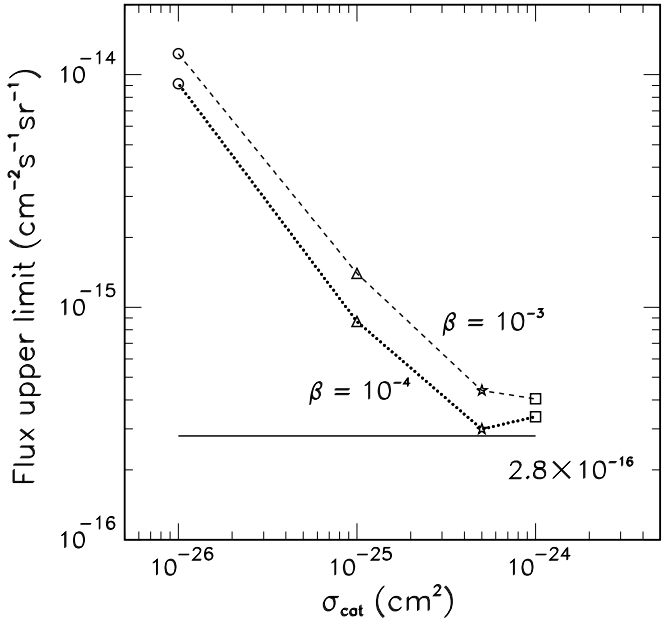}
}
\caption{Upper limits to the MM flux as a function of the catalysis cross
section for two different MM velocities, as determined by this analysis.}
\label{limsig}
\end{figure}

Fig.\ \ref{limsig} shows how the upper limit to the MM flux obtained with
this new catalysis-oriented analysis varies as a function of the catalysis
cross section $\sigma_{cat}$ for two MM velocities $\beta = 10^{-4}$ and
$\beta = 10^{-3}$\/.

\section{Discussion of the results}
As it can be seen from Fig.\ \ref{limits}, the MM flux upper limit is
very good at lower $\beta$, and becomes worse as the MM velocity
increases. This trend clearly reflects the analysis efficiency behaviour.
Moreover there quite no difference between the two theoretical models on the
proton cross section form.

From Fig.\ \ref{limsig} we see that the upper limits are very stringent for
higher cross sections, while for lower values, $\sigma_{cat} \Ltsim
10^{-25}~cm^2$ they are not very significant. This holds for all MM
velocities.

In Table 2 the results obtained by some experiments searching for MM
induced nucleon decay are summarized, together with the MACRO upper limit. It
should be noted that, except partly for Soudan1, the limits obtained in this way
rely on the catalysis process, so bounds to the MM flux set by these
experiments depend on the catalysis cross section: if the catalysis process
does not take place there are no limits at all. Moreover the underwater and
Cherenkov detectors must assume a very high catalysis cross section to attain
the significant upper limits they quote. On the other hand MACRO can set
significant limits to the MM flux, at the same level or even better than
those experiments, with fewer assumptions, and its limits stand more or less
the same even if the catalysis process does not take place or has a negligible
cross section. In comparison with other experiments MACRO is very sensitive to
small cross sections and can give good upper limits even in the presence of
higher values of $\sigma_{cat}$\/.

\begin{table*}[htb]
\begin{center}
\caption{Magnetic monopole flux limits (in $cm^{-2} s^{-1} sr^{-1}$) obtained by
experiments searching for nucleon decays catalyzed by a GUT MM}
\begin{tabular}{ccccc}
\hline
Experiment & Technique & Flux limit & $\beta$ range \\
\hline
Soudan1\cite{bartelt87} & Proportional tubes &
$8.8\cdot 10^{-14}$ & $10^{-2} \div 1$ \\
IMB\cite{becker94} & Water Cherenkov &
$1\div 3\cdot 10^{-15} $ & $10^{-5} \div 10^{-1}$ \\
Kamiokande\cite{kajita85} & Water Cherenkov &
$2.5\cdot 10^{-15}$ & $5\cdot 10^{-5} \div 10^{-3}$ \\
Baikal\cite{balkanov98} & Underwater detector &
$6\cdot 10^{-17}$ & $\simeq 10^{-5}$ \\
\hline
MACRO & Streamer tubes &
$3\cdot 10^{-16}$ & $1.1\cdot 10^{-4} \div 5\cdot 10^{-3}$ \\
\hline
\end{tabular}
\end{center}
\end{table*}

Indirect bounds to the MM flux can be set also by measuring the energy
emission from astrophysical objects like neutron stars and white dwarfs, which
could capture and accumulate MMs in their interior: these MMs could
then catalyze the decay of the nuclear matter inside them. The best limit is
obtained from the measure of the diffuse X-ray background, which essentially
the oldest neutron stars contribute to, and which is of the order of
$10^{-23}~cm^{-2}sr^{-1}s^{-1}$ \cite{kolb84}\/ (for $\beta = 10^{-3}$ and
$\sigma = 10^{-25}~cm^2$); limits can be derived from the emission of single
massive objects, such as the white dwarf WD 1136-286, at the level of
$10^{-20}~cm^{-2}sr^{-1}s^{-1}$ \cite{freese99}\/(again for $\beta = 10^{-3}$
and $\sigma = 10^{-25}~cm^2$). Although very impressive, nevertheless these
limits are indirect and very model--dependent, especially for what concerns the
estimate of the density of old neutron stars, based on the assumed birth
rate, the not well known X-ray absorption in the interstellar medium, and the
uncertainties in the source distance measurements. There could be also physical
loopholes to the validity of these limits \cite{kolb84}. Of course if the
MMs cannot catalyze the nucleon decay there are no bounds at all.
\begin{figure}
\resizebox{0.5\textwidth}{!}{
  \includegraphics{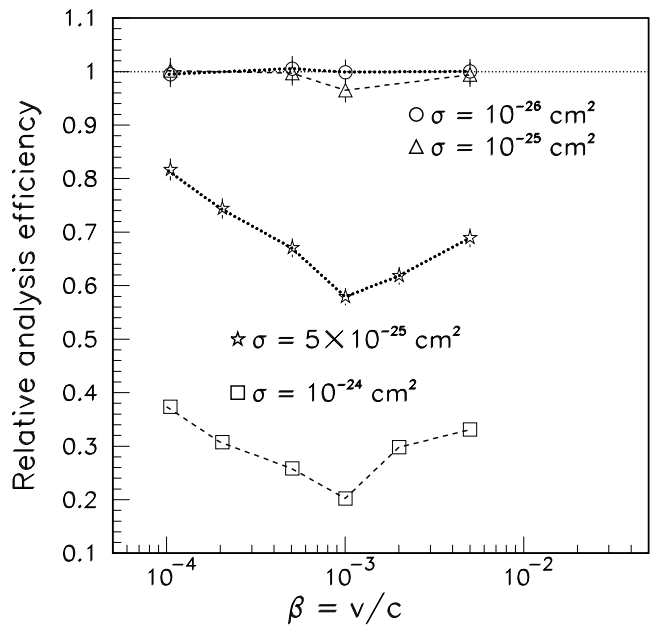}
}
\caption{Relative efficiency of the standard direct analysis applied to the
simulated samples with catalysis (using the model with equal cross sections)
with respect to the simulated sample without catalysis.}
\label{standrat}
\end{figure}

\section{Effects of the catalysis on the standard streamer tube analysis}
The standard streamer tube data analysis (along with the related hardware) is
described in detail in Ref.\ \cite{ambrosio95}\/. Briefly it is based on the
search for single tracks in space and time, with at least 7 points in the wire
and time views and at least 6 points in the strip view; all good tracks are
required to have $|\beta| \le 5\cdot 10^{-3}$\/. As in the catalysis search,
only the horizontal streamer planes of the lower MACRO participated in the
trigger formation, both lower and upper streamer planes were used for the event
reconstruction.

The standard MM search with the streamer tubes assumes that the MM
does not induce the nucleon decay, hence its validity for $\sigma_{cat} \Ltsim
10^{-27}~cm^2 = 1~mb$\/. To study the dependency of this analysis on the
catalysis cross section, all the simulated samples with catalysis were analyzed
with exactly the same programs used for the real data, applying all but the cut
on $\beta$\/, and the results were compared with the simulated sample without
catalysis. In Fig.\ \ref{standrat} the ratio of the analysis efficiency for the
samples with a given cross section with respect to the analysis efficiency
for the sample with $\sigma_{cat} = 0$ is plotted. The behaviour of this ratio
as a function of the total cross section is clear: for $\sigma_{cat} =
10^{-26}~cm^2 = 10~mb$ the analysis efficiency is actually unchanged with
respect to the case of no catalysis; also for $\sigma_{cat} = 10^{-25}~cm^2 =
100~mb$ the efficiency is quite the same. So in both cases the catalysis does
not decrease the MACRO capabilities of detecting a MM, and can even help
its detection, since there is an increase in the number of triggering events
(Fig.\ \ref{triggers}). Instead for $\sigma_{cat} = 10^{-24}~cm^2 = 1000~mb$
the events are so complex and rich of hits that the standard analysis does not
succeed in reconstructing most of them. This is due to the fact that at this
level there are so many catalysis hits that the reconstruction procedure may
fail. We think that $1000~mb$ is the maximum value worth to be considered, any
simulation for higher values being meaningless. From this point of view
$\sigma_{cat} = 5\cdot 10^{-25}~cm^2 = 500~mb$ can be considered an
intermediate case: the typical event is enough complicated to put in trouble
the reconstruction code, but not so much as to defeat it completely.

In addition, the behaviour of the analysis efficiency as a function of the
MM $\beta$ is interesting: in all samples this efficiency is minimum for
intermediate velocities, particularly for $\beta = 10^{-3}$\/. At low
velocities the time track is very inclined, since the MM ToF is large, so
the fit procedure can more easily distinguish it among the catalysis hits. At
high velocities the fit may mix the MM hits and the catalysis hits, but
since the time track is near vertical this does not imply a significant error
in the ToF reconstruction. Instead, at intermediate velocities this hit mixing
is very deleterious leading often to a wrong ToF estimation. In this regime in
fact many of the rejected events have a reconstructed $\beta$ much greater than
the actual one, often near to that of a relativistic particle. Clearly, this
effect is more evident for higher $\sigma_{cat}$ values.
\begin{figure*}
\resizebox{\textwidth}{!}{
  \includegraphics{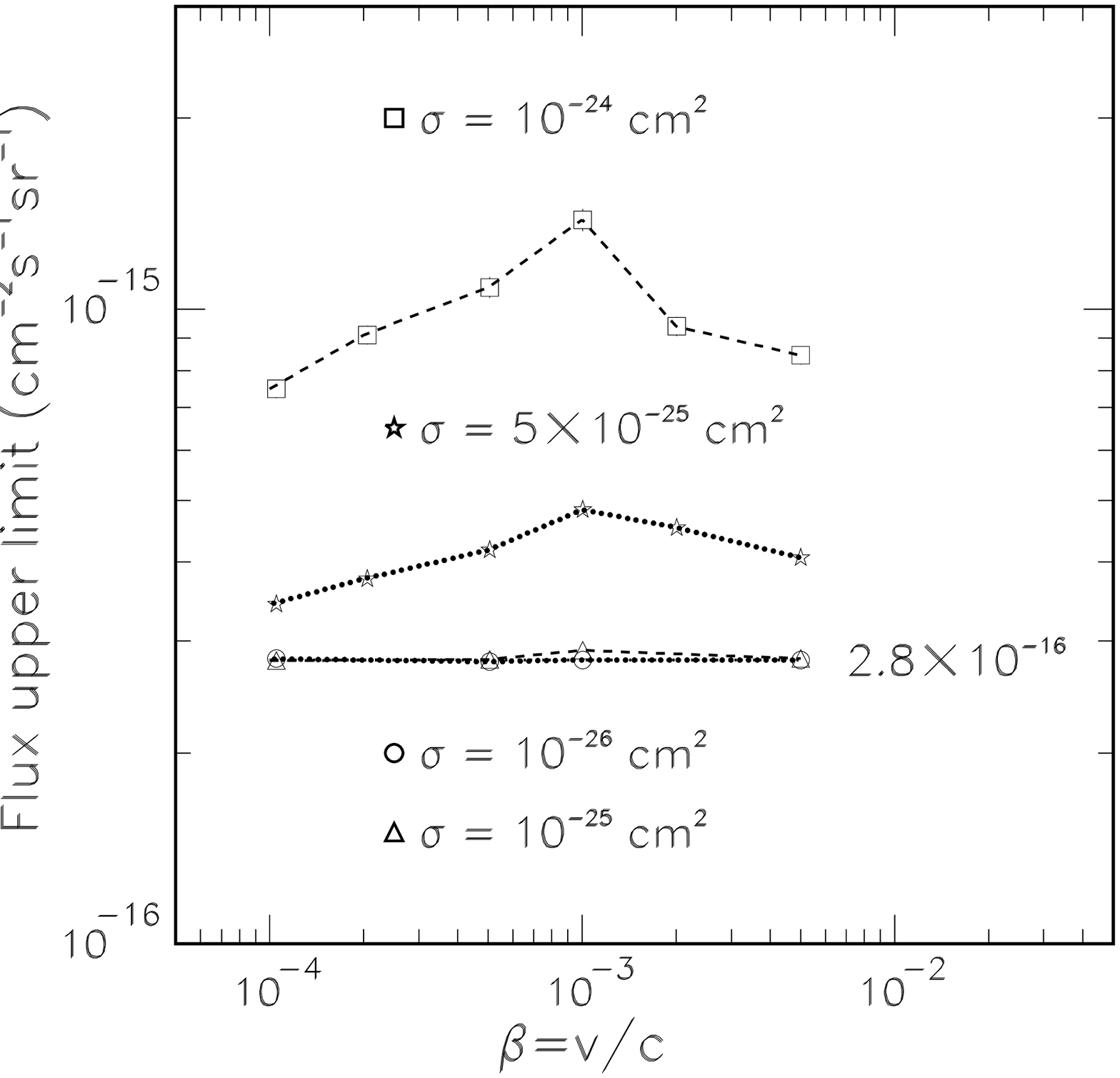}
  \includegraphics{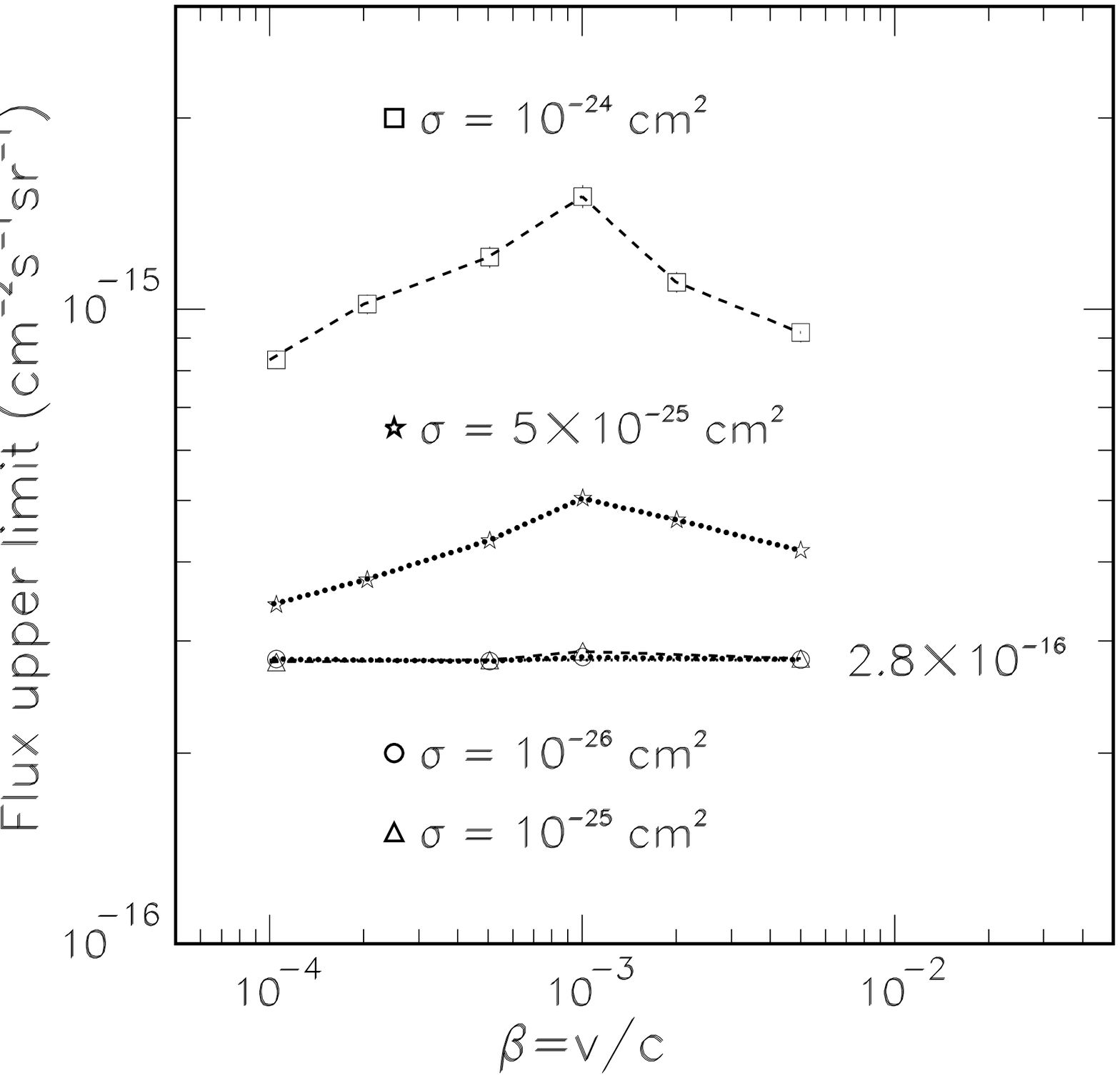}
}
\caption{Upper limits to the MM flux as a function of the MM
velocity for various catalysis cross section, as determined by the standard
analysis taking into account also the catalysis process, assuming the two
theoretical models for the cross section form, equal for proton and neutron
(left) or proton enhanced (right).}
\label{standlim}
\end{figure*}

Since the data set is exactly the same as the one used in the standard
published analysis, we can use the same acceptance and live time, while the
analysis efficiency must be corrected for the catalysis effects. In Fig.\
\ref{standlim} we show the new upper limits to the MM flux as obtained
with the standard analysis taking into account the catalysis process as well
(the quoted number is the published limit neglecting this process). For
$\sigma_{cat} = 5\cdot 10^{-25}~cm^2$ the new bounds are between $3.43$ and
$4.83\cdot 10^{-16}~cm^{-2}~s^{-1}~sr^{-1}$ for the model assuming the same
decay cross section for proton and neutron, and between $3.42$ and $5.04\cdot
10^{-16} cm^{-2} s^{-1}$ $sr^{-1}$ in the model with enhanced proton decay
cross section.

\begin{figure}
\resizebox{0.5\textwidth}{!}{
  \includegraphics{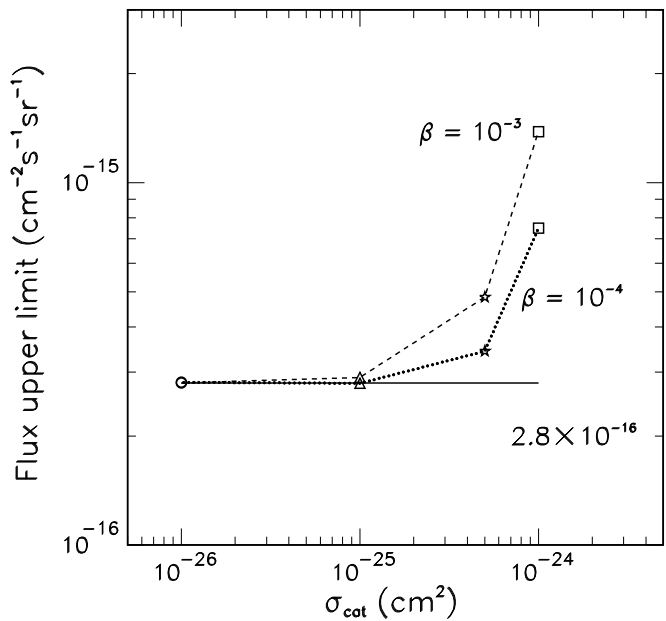}
}
\caption{Upper limits to the MM flux as a function of the catalysis cross
section for two different MM velocities, as determined by the previously
published direct analysis but taking into account the catalysis process as
well.}
\label{standsig}
\end{figure}

Fig.\ \ref{standsig} shows how the MACRO upper limit to the MM flux
obtained with the previously published ana\-lysis changes as a function of the
catalysis cross section $\sigma_{cat}$ for two MM velocities $\beta =
10^{-4}$ and $\beta = 10^{-3}$\/.

From the last Figure it can be clearly seen that the MM upper limit from
the published standard streamer analysis is practically unspoiled by the
catalysis process for lower cross section values. So the results of the
standard streamer MM analysis, previously reported for $\sigma_{cat} <
1~mb$ \cite{monofin}\/, can be extended to higher values of the catalysis cross
section up to $\sigma_{cat} \simeq 100~mb$\/. Even assuming a cross section
$\sigma_{cat} \simeq 500~mb$\/, the flux limit would not increase more than a
factor 2.

\section{Conclusions}
In this paper we presented the results of a dedicated search for events of
nucleon decays catalyzed by the passage of a GUT MM in the MACRO
detector. The efficiency of the present catalysis--oriented analysis, which
used only the streamer tube subdetector, is very good for catalysis cross
sections of the order $100\div 1000~mb$\/. Since no candidate survived, a 90\%
C.L.\ upper limit to the MM flux can be set at the level of $\sim 3\cdot
10^{-16}~cm^{-2}~s^{-1}~sr^{-1}$ in the range $10^{-4} \le \beta \le 5\cdot
10^{-3}$\/.

We have shown that a dedicated analysis based on the nucleon decay catalysis
yields flux limits which are considerably more stringent than those obtained by
nucleon decay experiments; we also made looser assumptions (smaller, more
reasonable cross sections, no need to have a chain of decay events to clearly
identify the MM). We were able to set upper limits to the local MM
flux even if the catalysis process does not take place, using a different
analysis strategy; here we have shown how this previous analysis, developed in
the hypothesis of no decay, depends on the catalysis cross section.

\vskip 1.cm
\section*{Acknowledgements}
We gratefully acknowledge the support of the director and of the staff of the
Laboratori Nazionali del Gran Sasso and the invaluable assistance of the
technical staff of the Institutions participating in the experiment. We thank
the Istituto Nazionale di Fisica Nucleare (INFN), the U.S. Department of Energy
and the U.S. National Science Foundation for their generous support of the
MACRO experiment. We thank INFN, ICTP (Trieste), WorldLab and NATO for
providing fellowships and grants (FAI) for non Italian citizens.

\end{document}